\documentclass[manuscript,screen]{acmart}

\AtBeginDocument{%
  \providecommand\BibTeX{{%
    \normalfont B\kern-0.5em{\scshape i\kern-0.25em b}\kern-0.8em\TeX}}}

\setcopyright{rightsretained}
\acmJournal{TECS}
\acmYear{2023} \acmVolume{} \acmNumber{} \acmArticle{} \acmMonth{1} \acmPrice{}\acmDOI{10.1145/3579644}

\usepackage{amsmath,amsfonts} 
\usepackage{algorithm}
\usepackage{algorithmic}
\usepackage{graphicx}
\usepackage{textcomp}
\usepackage{xcolor}

\usepackage{courier}
\usepackage{enumerate}
\usepackage{stmaryrd}
\usepackage{subfigure}

\usepackage{moresize}

\usepackage[backgroundcolor=black,textcolor=white,textsize=tiny]{todonotes}

\usepackage{listings}
\usepackage{multirow}
\usepackage[all,cmtip]{xy}
\usepackage{tikz}
\usetikzlibrary{automata, positioning, arrows.meta, decorations.markings}
\definecolor{improve}{gray}{0.5}
\definecolor{tikz.lightgrey}{rgb}{0.6, 0.6, 0.6}
\definecolor{tikz.red}{RGB}{190, 30, 45}
\definecolor{tikz.green}{RGB}{0, 147, 68}
\definecolor{tikz.blue}{RGB}{27, 117, 187}
\definecolor{tikz.darkblue}{RGB}{14, 69, 88}
\definecolor{tikz.yellow}{RGB}{251, 175, 63}
\tikzset{
    ->,  
    >=stealth, 
    node distance=1cm, 
    every state/.style={
        inner sep=2pt,
        minimum size=0.25cm,
    }, 
    initial text=$ $, 
    font=\tiny\ttfamily
}

\def\BibTeX{{\rm B\kern-.05em{\sc i\kern-.025em b}\kern-.08em
    T\kern-.1667em\lower.7ex\hbox{E}\kern-.125emX}}

\usepackage{eso-pic}

\begin{document}

\AddToShipoutPictureBG*{%
\AtPageUpperLeft{%
\setlength\unitlength{1in}%
\hspace*{\dimexpr0.5\paperwidth\relax}
\makebox(0,-1.5)[c]{
\begin{tabular}{c c}
Sander Thuijsman and Michel Reniers,
Supervisory Control for Dynamic Feature Configuration in Product Lines, \\
Accepted for: {\em ACM Transactions on Embedded Computing Systems} (2023)\\
Available open access: {\url{https://www.doi.org/10.1145/3579644}}\\
Uploaded to ArXiv \today. \\
\end{tabular}}}}

\title{Supervisory Control for Dynamic Feature Configuration in Product Lines}

\author{Sander Thuijsman}
\email{s.b.thuijsman@tue.nl}
\orcid{0000-0002-1628-8622}
\author{Michel Reniers}
\email{m.a.reniers@tue.nl}
\orcid{0000-0002-9283-4074}
\affiliation{%
  \institution{Eindhoven University of Technology}
  \streetaddress{P.O. Box 513}
  \city{Eindhoven}
  \country{The Netherlands}
  \postcode{5600~MB}
}


\begin{abstract}
In this paper a framework for engineering supervisory controllers for product lines with dynamic feature configuration is proposed. 
The variability in valid configurations is described by a feature model. 
Behavior of system components is achieved using (extended) finite automata and both behavioral and dynamic configuration constraints are expressed by means of requirements as is common in supervisory control theory.
Supervisory controller synthesis is applied to compute a behavioral model in which the requirements are adhered to. 
For the challenges that arise in this setting, multiple solutions are discussed. 
The solutions are exemplified in the CIF toolset using a model of a coffee machine.
A use case of the much larger Body Comfort System product line is performed to showcase feasibility for industrial-sized systems.
\end{abstract}




\maketitle

\section{Introduction}
\label{sec:introduction}
In present day development of systems and products, reuse of both software and hardware components is sought to reduce development and production costs, and shorten time-to-market. 
The goal of Software/System Product Line Engineering (SPLE) is to facilitate reuse throughout all phases of systems engineering \cite{Pohl2005}.
Adoption of this paradigm requires identification of the core assets of the products in the domain in order to exploit their commonality and manage their variability, often defined in terms of features.
A feature is defined as a logical unit of behavior specified by a set of functional and non-functional requirements \cite{Bosch} or a distinguishable characteristic of a concept (system, component, etc.) that is relevant to some stakeholder \cite{Czarnecki}. 
Feature models may be used to define which combinations of features are considered valid product configurations \cite{Benavides2010}.

In literature there has been much attention for correct configuration of SPLs \cite{Benavides2010}. 
Behavioral correctness is studied only recently, since \cite{Classen2010}.
Typically the approaches that are used for guaranteeing a proper functioning SPL (i.e., correct with respect to its requirements or specifications) are verification technologies such as theorem provers \cite{Classen2013}, model checkers \cite{Baier}, and correct-by-construction approaches such as supervisory controller synthesis \cite{Beek2016}. 
In \cite{Beek2016}, for the first time supervisory controller synthesis \cite{Ramadge1984,Ramadge1987} has been considered for constructing supervisory controllers for an SPL described by a feature model.

Supervisory control theory, as introduced by \cite{Ramadge1987}, is a model-based approach to control discrete event systems.
In this framework a model is created of the uncontrolled system, and behavioral requirements are specified that define what behavior is allowed.
Using these models, a supervisory controller can be computed algorithmically (synthesized), such that it restricts the behavior of the system to always be in accordance with the requirements.
Depending on the synthesis algorithm, the behavior of the system under control is guaranteed to have some useful properties, such as safety, nonblockingness, controllability, and maximal permissiveness. 
The benefits of supervisory control theory have been demonstrated in industrial use cases, such as for example supervisory control of lithography machines \cite{VanDerSanden2015}, health-care systems \cite{Theunissen2014}, automotive applications \cite{KorssenDMRH18}, and infrastructural systems \cite{Reijnen2020}. 
Typically, the models that are input to supervisory controller synthesis are discrete-event system models such as (extended) finite automata \cite{Cassandras2008,SkoldstamAF07}.

The contribution of this work is a model-based framework for the supervisory control of product lines.
This approach consists out of the following steps:
\begin{enumerate}
    \item \textit{Representing the feature model in extended finite automata.} This is prerequisite, because we apply supervisory controller synthesis that is based on automata specifications.
    \item \textit{Capturing dynamic configuration of features in the models.} In this work, we pay additional attention to the situation where features might enter or leave the system during runtime.
    \item \textit{Modeling uncontrolled system behavior such that it properly takes the current configuration into account.} A component-wise specification of the system behavior is given, where the component behavior is linked to the presence of features in the configuration.
    \item \textit{Modeling behavioral requirements depending on presence of features.} The requirements of the behavior are dependent on the current configuration. Additionally, different requirements may apply when the system is in a transitional phase in between valid configurations.
    \item \textit{Applying supervisory controller synthesis.} A correct-by-construction supervisory controller is obtained from the developed models.
\end{enumerate}

For most of these steps, there are multiple solutions and it depends on the case at hand which one is most appropriate. 
We mention the alternatives and illustrate them.
To exemplify the method discussed in this paper, we use the coffee machine system from \cite{Beek2014} as a running example. 
Modeling of automata and supervisory controller synthesis is performed using the tool CIF \cite{VanBeek2014}.
Scalability and applicability of the approach is later demonstrated using the Body Comfort System (BCS) from \cite{Lity2013}.

\subsection{Related work}
This work is based on, and can be seen as an extension to, \cite{Beek2016} and \cite{Thuijsman2020}. 
The basis of the approach we present here was first introduced in \cite{Beek2016}, where feature models are modeled in CIF, behavioral models of the system components are defined, and a supervisory controller is obtained that takes into account the possible configurations as defined by the feature model.
In \cite{Thuijsman2020}, this work was extended by also considering the setting of dynamic configuration, where components are allowed to enter and leave the system.
Relative to \cite{Beek2016,Thuijsman2020}, the extension we present here includes more modeling possibilities, considerations, explanations, and examples. 
Additionally, this work also provides a case study of the large BCS use case, showcasing applicability for industrial-sized product line systems.

Below we mention some more related work, which we divide into the following categories: (1) works that study dynamic reconfiguration during run-time, but do not apply supervisory control theory, (2) works that apply supervisory control theory for multiple configurations during design-time, (3) works that apply supervisory control for dynamic reconfiguration during run-time.
Our work fits the latest category. 
However, our work differentiates from the mentioned works, as in none of them dynamic feature configuration in relation to supervisory control engineering with a clear separation of uncontrolled system behavior and specification of behavioral and dynamic reconfiguration requirements is discussed.

\subsubsection{Dynamic configuration during run-time}

In \cite{Kogekar2004} an approach for dynamic software reconfiguration in sensor networks is presented. The dynamic reconfiguration is based on formal constraints in terms of quality-of-service parameters that are measured at runtime. 

Dynamic runtime variability of software product lines in embedded automotive software systems is applied to create
adaptable and reconfigurable software architectures in \cite{Shokry2008}.
Also \cite{Rosenmuller2011} discusses reconfiguration with the purpose of determining an optimal configuration at runtime. 
In both papers the dynamic configuration is under control, which is typically not the case in the present paper.

In \cite{Shen2011}, a feature-oriented method is proposed to support runtime variability reconfiguration by introducing an intermediate level between feature variations and implementations.

In \cite{Gharsellaoui2017}, the authors deal with reconfiguration of real-time embedded systems to cope with hardware/software faults.

The authors of \cite{Sharifloo2016} argue that it is not reasonable to anticipate all relevant context changes during design-time and therefore propose a model that combines learning of adaptation rules with evolution of the configuration space, which can be applied during run-time.

\subsubsection{Supervisory control and design-time system configuration}

In \cite{Thuijsman2022} the assumption is made of no a-priori knowledge of the possible system configurations.
Computation of a supervisory controller for an updated system, given knowledge over the base system, is studied.

In \cite{Basile2019}, priced featured automata were translated to extended finite automata and the structure of the SPL was used to greatly reduce the number of supervisory controller syntheses required to solve game-based energy problems.

In \cite{Kahraman2021} feature models are used to generated product instances for the model-based engineering tool LSAT. 
LSAT is a tool used to design supervisory controllers, but it can not perform supervisory controller synthesis \cite{vanderSanden2021}.

In \cite{Verbakel2021} a method to obtain supervisory controllers for a product family is discussed through the use of a configurator, where synthesis is applied after selection of parameterized components.

\subsubsection{Supervisory control and dynamic configuration during run-time}

\cite{Basile2020} apply supervisor synthesis to featured modal contract automata.
They synthesize orchestrations, that match service requests to service offers, for all valid products in a product line, by joining the orchestrations of a small subset of the valid products.
By means of a composition operation, the product line can dynamically be updated and new services can join composite services.

\subsection{Structure}
In Section \ref{sec:preliminaries} we introduce feature models and the CIF language. 
Using the CIF language, we show how feature models can be represented in automata format in Section \ref{sec:modeling_feature_models}.
In Section \ref{sec:dynamic}, the use of automata to represent dynamically configured feature models is discussed. 
Modeling of component behavior in the setting of dynamic configuration is discussed in Section \ref{sec:modeling_behavior}.
In Section \ref{sec:requirements}, the modeling of behavioral requirements that are dependent on the current configuration is discussed.
Supervisory controller synthesis is applied in Section \ref{sec:supsynthesis}.
An industrial-sized use case is discussed in Section \ref{sec:usecase}.
Section \ref{sec:conclusions} concludes the paper. 

\section{Preliminaries}
\label{sec:preliminaries}

\subsection{Feature Models}
A feature model \cite{Heymans2008} is a graph with a collection of nodes representing features, and a number of relations between these features, called feature constraints. 
The feature constraints that can be expressed are summarized in Table \ref{table:relations}, which is taken from \cite{Beek2016}. 
The right column provides logical formulas that express how presence of the features (denoted by the $F_i$) is restricted by the different types of constraints.

For any valid configuration a \textit{root} feature needs to be present. 
\textit{Mandatory} features are required to be present when their parents are, and optional features may be present when their parents are.
For a set of \textit{alternative} features, exactly one is present when their parent is present. 
And for a set of \textit{or} features, at least one is present when their parent is present.
It can also be defined that the presence of a certain feature \textit{requires} or \textit{excludes} another feature to be present.

\begin{table}[htpb]
\setlength{\tabcolsep}{15pt}
\renewcommand{\arraystretch}{0.9}
\caption{Different feature constraints of a feature model \cite{Beek2016}.}
\label{table:relations}
\begin{center}
\begin{tabular}{|c|c|c|}
\hline
\multicolumn{2}{|c|}{Constraint} & Formula\rule[-4.5pt]{0pt}{15pt} \\
\hline
\hline
\multirow{2}{*}{root}
&
\multirow{2}{*}{$\def\objectstyle{\scriptstyle}\def\labelstyle{\scriptstyle}
\entrymodifiers={+[F][F-]}\xymatrix@C=.25cm@R=.25cm{
*{} & {F_0} & *{} \\
}$}
&
\multirow{2}{*}{$F_0\iff\mathit{true\/}$} \\
& & \\
\hline
\multirow{4}{*}{mandatory}
&
\multirow{4}{*}{$\def\objectstyle{\scriptstyle}\def\labelstyle{\scriptstyle}
\entrymodifiers={+[F][F-]}\xymatrix@C=.25cm@R=.25cm{
*{} & {F_1}\ar@{-}[dd]+(0,2) & *{} \\ 
*{} & *{} & *{} \\
*{} & {F_2} & *{} \\
}$}
&
\multirow{4}{*}{$F_1\iff F_2$} \\
& & \\ & & \\  & & \\
\hline
\multirow{4}{*}{optional}
&
\multirow{4}{*}{$\def\objectstyle{\scriptstyle}\def\labelstyle{\scriptstyle}
\entrymodifiers={+[F][F-]}\xymatrix@C=.25cm@R=.25cm{
*{} & {F_1}\ar@{-o}[dd]+(0,2.5) & *{} \\
*{} & *{} & *{} \\
*{} & {F_2} & *{} \\
}$}
&
\multirow{4}{*}{$F_2\implies F_1$} \\
& & \\ & & \\ & & \\
\hline
\multirow{4}{*}{alternative}
&
\multirow{4}{*}{$\def\objectstyle{\scriptstyle}\def\labelstyle{\scriptstyle}
\entrymodifiers={+[F][F-]}\xymatrix@C=.25cm@R=.25cm{
*{} & *{} & {F} \ar@{-}[dd]|{\rule{.8cm}{.5pt}} \POS[]!DC;[ddl]!UC**\dir{-}, []!DC;[ddr]!UC**\dir{-} & *{} & *{} \\
*{} & *{} & *{} & *{} & *{} \\ 
*{} & {F_1} & {F_2} & {F_n} & *{} \\
}$}
&
\multirow{2}{*}{$(F_1\!\iff\!(\neg F_2\land\cdots\land \neg F_n\land F))$\quad} \\
& & \multirow{2}{*}{$\;\;\;{}\land\cdots\land{}$} \\ 
& & \multirow{2}{*}{$(F_n\!\iff\!(\neg F_1\hspace*{0.1em}\land\cdots\land \neg F_{n-1}\land F))$} \\
& & \\
\hline
\multirow{4}{*}{or}
&
\multirow{4}{*}{$\def\objectstyle{\scriptstyle}\def\labelstyle{\scriptstyle}
\entrymodifiers={+[F][F-]}\xymatrix@C=.25cm@R=.25cm{
*{} & *{} & {F} \ar@{-}[dd]|{
\raisebox{.25cm}{\begin{tikzpicture}[scale=.25] \draw[fill=black] (0,0) --  (3,0) -- (1.5,1) -- cycle; \end{tikzpicture}}
} \POS[]!DC;[ddl]!UC**\dir{-}, \POS[]!DC;[dd]!UC**\dir{-}, []!DC;[ddr]!UC**\dir{-} & *{} & *{} \\
*{} & *{} & *{} & *{} & *{} \\
*{} & {F_1} & {F_2} & {F_n} & *{} \\
}$}
&
\multirow{4}{*}{$F\iff(F_1\lor F_2\lor\cdots\lor F_n)$} \\
& & \\ & & \\ & & \\
\hline
\multirow{2}{*}{requires}
&
\multirow{2}{*}{$\def\objectstyle{\scriptstyle}\def\labelstyle{\scriptstyle}
\entrymodifiers={+[F][F-]}\xymatrix@C=.25cm@R=.25cm{
*{} & {F_1}\ar@{{}*{\txt{\raisebox{-1.8mm}{-}}}>}[rrr] 
& *{} & *{} & {F_2} & *{} \\
}$}
&
\multirow{2}{*}{$F_1\implies F_2$} \\
& & \\
\hline
\multirow{2}{*}{excludes}
&
\multirow{2}{*}{$\def\objectstyle{\scriptstyle}\def\labelstyle{\scriptstyle}
\entrymodifiers={+[F][F-]}\xymatrix@C=.25cm@R=.25cm{
*{} & {F_1}\ar@{<*{\txt{\raisebox{-1.8mm}{-}}}>}[rrr] 
& *{} & *{} & {F_2} & *{} \\
}$}
&
\multirow{2}{*}{$\neg\,(F_1\land F_2)$} \\
& & \\
\hline
\end{tabular}
\end{center}
\end{table}

In an \textit{extended} feature model \cite{Benavides05}, attributes can be assigned to features.
An example of such an attribute could be the weight or price of a feature.
Attributes are typically defined by a name, domain (such as integers, enumerations, etc.), and a value.
Attribute constraints can be expressed using attributes of features.  
A constraint could be a maximal value for the total weight or price of the system. 
We may use the term \textit{feature model} to refer to both `normal'  and extended feature models when the type of feature model is not relevant or clear from the context.




\begin{example}[Feature model for a coffee machine]
Consider a product line for a coffee machine \cite{Beek2014}.
An extended feature model that captures the allowed configurations for this product line is presented in Fig. \ref{fig:CoffeeMachine_FM}. 
In the solid boxes, the features' names are shown with an abbreviation. 
We observe that the coffee machine always contains a sweet, coin, and beverage feature. 
Optionally, the machine may also be able to sound a ringtone or return change.
The machine accepts euro or alternatively dollar coins.
The machine always offers coffee as a beverage, but may optionally also offer cappuccino or tea.
If the machine offers cappuccino, it can not accept dollar coins.
When the machine offers cappuccino, it is required the machine comes equipped with the ringtone feature.

Fig. \ref{fig:CoffeeMachine_FM} also shows dashed boxes, in which the feature attributes are shown with a name, domain, and value.
Some features have no attribute, and some features have a cost attribute.
The cost is valued by an integer, and the cost values are between 3 and 10.
The total machine cost is the sum of costs of all features that have a cost attribute.
Using the attributes, one can formulate an attribute constraint, e.g., the total machine cost must be less than or equal to 30.

The feature model represents 20 valid configurations.
If we include the previously mentioned cost constraint, there are 16 valid configurations.

\begin{figure}[thpb]
\centerline{
\includegraphics[width=0.8\columnwidth]{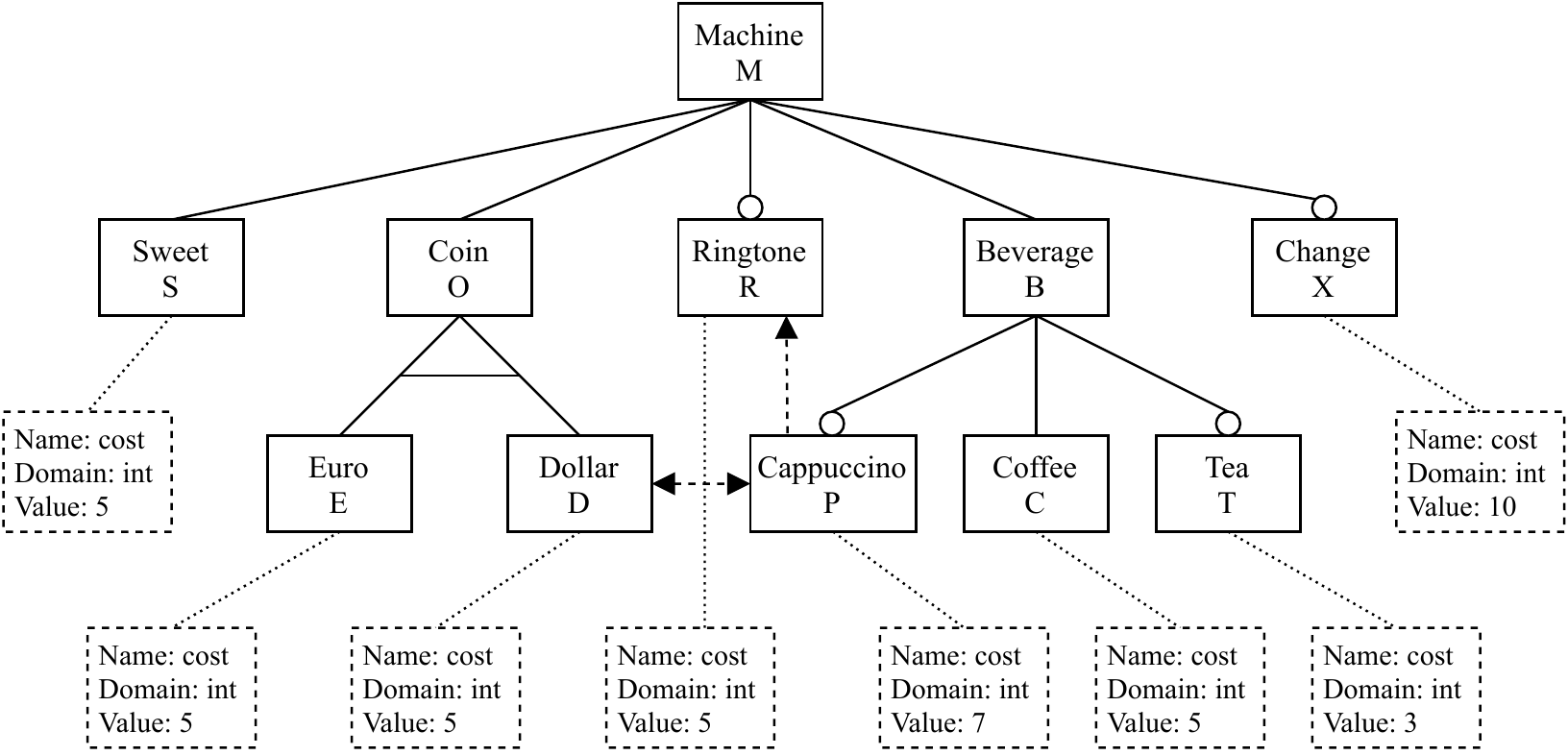}
}
\caption{Feature model of the coffee machine \cite{Beek2014}.}
\label{fig:CoffeeMachine_FM}
\end{figure}

\end{example}

\subsection{CIF}
There are two tool suites that support supervisory controller synthesis for models expressed as extended finite automata: Supremica \cite{Supremica} and CIF \cite{VanBeek2014}. 
In \cite{Beek2016}, it has been shown how the CIF language and tool set can be used for synthesizing a supervisory controller that is suited for an SPL. 
The approach uses the concept of algebraic variables extensively, which is not available in Supremica.

CIF, part of the Eclipse Supervisory Control Engineering Toolkit (Eclipse ESCET™)\footnote{The ESCET toolset and documentation is open source and freely available at \url{https://www.eclipse.org/escet/}. `Eclipse', ‘Eclipse ESCET’ and ‘ESCET’ are trademarks of Eclipse Foundation, Inc.}, is a language and tool set that supports model-based engineering of supervisory controllers involving modeling, (visualized) simulation, synthesis, verification, and code generation \cite{VanBeek2014}.
In the past years CIF has been applied to many industrial-size case studies \cite{VanDerSanden2015,Theunissen2014,KorssenDMRH18,Reijnen2020}.
Although CIF allows modeling of real-valued variables that evolve continuously over time (as described by differential equations), for the purpose of this paper our attention is restricted to discrete-event models.

Discrete-event models of the uncontrolled system (also called plant) can be developed in the form of a collection of \emph{extended finite automata} \cite{SkoldstamAF07}. 
The automata that comprise the plant synchronize over shared events and interact through the reading of each others' (discrete) variables, and thus the global system behavior is achieved through synchronous composition \cite{Cassandras2008}.
An automaton consists of locations and edges between these locations. 
The edges are labeled by an event, a guard, and an update. 
The guard describes a condition (in terms of the variables) that enables the occurrence of the event associated with the edge. 
The update describes how the values of the variables change in such a transition. 
In CIF variables are declared inside an automaton and follow the `global read, local write' principle, which means that each variable may be inspected in any of the automata, but may only be adapted in its defining automaton.
A CIF automaton has at least one initial location, and variables have at least one initial value.
The \textit{state} of an automaton is defined by its current location and variable valuations.
Initial/marked locations implicitly define initial and marked states.

Events are defined to be \emph{controllable} or \emph{uncontrollable}. 
Uncontrollable events cannot be prevented from occurring by a supervisory controller, whereas controllable events can be blocked. 
The extended finite automata may have \emph{marked} states. 
Marked states are typically used to denote states in which the system has finished a task.
By applying supervisor controller synthesis, the controllable events are restricted in such a way that from every reachable state, a marked state can eventually be reached.

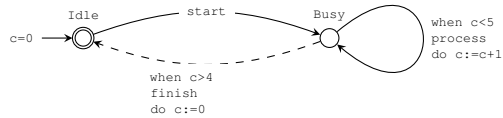
\begin{figure}[t]
\begin{lstlisting}[frame=single,caption={Textual CIF model of an automaton.},label={lst:example}]
plant automaton ExampleAutomaton:
controllable start, process;
uncontrollable finish;
disc int c = 0;
  location Idle: initial; marked;
    edge start goto Busy;
  location Busy:
    edge process when c<5 do c:=c+1;
    edge finish when c>4 do c:=0 goto Idle;
end
\end{lstlisting}
    \centering
    \begin{tikzpicture}
    \centering
        \node[state, initial, initial text = {c=0}, accepting,label={above:Idle}] (Idle) {};
        \node[state, right=3.0 of Idle,label={above:Busy}] (Busy) {};
        
    \draw 
    (Idle) edge[bend left=20] node[fill=white]{start} (Busy)
    (Busy) edge[bend left=20,dashed] node[fill=white,below]{\begin{minipage}[htpb]{1.5cm}when c>4\\
                finish\\
                do c:=0\end{minipage}} (Idle)
    (Busy) edge[loop, out=40,in=-40,looseness=30] node[right]{\begin{minipage}[htpb]{2cm}when c<5\\
        process\\
        do c:=c+1\end{minipage}} (Busy);
    \end{tikzpicture}
    \caption{Graphical representation of the automaton from Listing \ref{lst:example}.}
    \label{fig:example}
    \Description{figure description}
\end{figure}

An example of a CIF automaton is given in Listing \ref{lst:example}.
Its graphical representation is given in Fig. \ref{fig:example}. 
Locations are represented by small circles with their name next to them. 
Initial locations have a dangling incoming arrow and possibly an expression stating the initial values of the variables. 
In the example there is a \texttt{c} variable that is defined as a discrete (the value will update in a discrete manner) integer, denoted with the keywords \texttt{disc int}, which has an initial value of \texttt{0}.
Marked locations have a double circle representation. 
Edges that are labeled by a controllable event are represented by a solid arrow and edges with an uncontrollable event by a dashed arrow. 
The optional guard is indicated by the keyword \texttt{when} and the update using \texttt{do}.
In this paper we use both textual and graphical representations as we see fit.

In CIF requirements are specified by means of automata that state in which orderings the contained events are allowed to occur, or by using state-based expressions such as event conditions and state invariants \cite{MaWonham06,MarkovskiJBSR10}. 
An event condition restricts the occurrences of an event to situations where a certain condition in terms of the variables of the model is satisfied. 
A state invariant expresses in which states the system is allowed to be.

CIF has several concepts that facilitate modeling of large systems, such as a definition/instantiation mechanism for automata and requirements and algebraic variables. 
For an algebraic variable, the value is defined to be identical to the value of some expression (in terms of other variables). 
In this paper algebraic variables are used abundantly.

With each location $L$ in each automaton $A$, CIF associates a location variable $A.L$ that may be used in guards, in right-hand sides of updates and algebraic variables, and in state-based requirements. 
Updates of these variables are implicit and according to the location change of an automaton.

\section{Static feature models in CIF}
\label{sec:modeling_feature_models}
In this section we demonstrate modeling of feature models where features can not configure dynamically, i.e., they are static, as proposed in \cite{Beek2016}.
We first discuss normal (non-extended) feature models, and then attributes to allow extended feature models.
Next, in Section \ref{sec:dynamic} we will show how these static models can be extended to feature models supporting dynamic feature configuration, i..e., features may enter or leave the system during run-time.

A CIF model representing all allowed configurations for a given feature model is obtained as follows.
For each feature, an automaton is introduced that captures whether the feature is present or not. 
It uses a Boolean variable \texttt{present}, the value of which is fixed initially, that is \texttt{true} when the feature is present and \texttt{false} otherwise. 
This variable is also used to capture the feature constraints expressed in the feature model.

Since the feature automata have the same structure we use an automaton definition in CIF, which is then instantiated for each feature in the feature model. 
The CIF specification for this automaton definition is given in the first four lines of Listing \ref{lst:feature_definition_static}. 
In CIF, every automaton needs to have at least one location, hence the dummy location (without a name) defined in Listing~\ref{lst:feature_definition_static}. 
Note that the initial value of \texttt{present} is left implicit (is allowed to be either \texttt{true} or \texttt{false} by using the keywords \texttt{in any}).
For each feature in the feature model an instance of this feature automaton definition is obtained by a statement such as the ones in lines 6 and 7.
These instances act just like separately defined automata.
We can write \texttt{F1.present} to refer to the \texttt{present} variable in \texttt{F1}.

\begin{lstlisting}[frame=single,caption={Automaton definition for features and instantiation for features.\label{lst:feature_definition_static}}]
plant def FEATURE():
  disc bool present in any;
  location: initial; marked;
end

F1: FEATURE(); 
F2: FEATURE();
...
\end{lstlisting}

Feature constraints arising from a feature model can be modeled in CIF in such a way that the transformation from a feature model to a CIF model can easily be automated. 
For each of the constraint types in Table \ref{table:relations}, an algebraic CIF expression is shown in Listing \ref{lst:several_exp}. 
`\texttt{/$\!$/}' in the listing denotes that the remainder of the line is a comment, which we use to specify the constraint type.
Also more complex constraints between features can be formulated in this way.
These algebraic expressions can also be used to denote more complex constraints between features that are not considered here.

\begin{lstlisting}[frame=single,caption={Several feature constraint expressions.\label{lst:several_exp}}]
alg bool r1 = F0.present <=> true; %*{\color{gray}//root}*)
alg bool r2 = F1.present <=> F2.present; %*{\color{gray}//mandatory}*)
alg bool r3 = F2.present => F1.present; %*{\color{gray}//optional}*)
alg bool r4 = (F1.present <=> (not(F2.present) and F.present)) and (F2.present <=> (not(F1.present) and F.present)); %*{\color{gray}//alternative}*)
alg bool r5 = F.present <=> (F1.present or F2.present); %*{\color{gray}//or}*)
alg bool r6 = F1.present => F2.present; %*{\color{gray}//requires}*)
alg bool r7 = not (F1.present and F2.present); %*{\color{gray}//excludes}*)
\end{lstlisting}


A valid configuration is obtained if and only if all feature constraints are satisfied.
To this end, the algebraic expressions for the seperate feature constraints, such as in Listing \ref{lst:several_exp}, can be used.
In Listing \ref{lst:valid_configuration_static} we introduce an algebraic variable \texttt{sys\_valid} that evaluates to true if and only if all feature constraints are satisfied.
We also define automaton \texttt{Validity} in lines 3-5 in Listing \ref{lst:valid_configuration_static}. 
At the moment this automaton only states that the system initially is in a valid system configuration.

\begin{lstlisting}[frame=single,caption={Validity of configuration.\label{lst:valid_configuration_static}}]
alg bool sys_valid = r1 and r2 and r3 and ...;

plant automaton Validity:
  location: initial sys_valid; marked;
end
\end{lstlisting}
In an extended feature model, an attribute may be assigned to a feature. 
To model an attributed feature as an automaton, next to the presence feature one or more variables need to be declared to represent the attribute.
In Listing \ref{lst:attributedfeature} an example is given for a ball feature.
The ball feature is attributed with a color, that can either be red, yellow, or blue.
When the ball feature is not present, the color is not available (\texttt{NA}).
The domain of the color enumerator is defined in line 1 of Listing \ref{lst:attributedfeature}.
In line 5 the \texttt{color} variable is defined for the ball feature;
it is an algebraic variable that takes the value defined in the feature automaton instantiation (\texttt{clr}) when the feature is present, and is \texttt{NA} otherwise.

\begin{lstlisting}[frame=single,caption={Attributed ball feature.\label{lst:attributedfeature}}]
enum colordomain =  red, yellow, blue, NA;

plant def BallFeature(alg colordomain clr):
  disc bool present in any;
  alg colordomain color = if present : clr else NA end;
  location: initial; marked;
end

RedBall: BallFeature(red);
YellowBall: BallFeature(yellow);
\end{lstlisting}

\begin{example}[Static feature model for coffee machine in CIF]
\label{ex:static}
The CIF specification of the feature model for the coffee machine is given in Listing \ref{lst:featureinstances_ws}
\footnote{The CIF models used in this paper are available here: \url{https://github.com/sbthuijsman/TECS_PLE}}.
Two different plant definitions for features are used; one without any attributes, and one with a cost attribute.
The plant definition for the attributed feature can be instantiated with an input variable defining the cost value.
An algebraic integer variable \texttt{cost\_sum} is introduced in line 39, as the sum of costs of all (present) features.
Then, an algebraic Boolean is defined that is true when \texttt{cost\_sum} is 30 or less.
In the \texttt{Validity} automaton, it is defined that initially both the feature and cost constraints are satisfied.
Construction of the state space of this model in CIF results in a structure with 16 allowed configurations, each represented by an initial state (and nothing more as we have not yet modeled any behavior).


\begin{lstlisting}[frame=single,caption={Feature instances of the coffee machine.\label{lst:featureinstances_ws}}]
plant def FEATURE():
  disc bool present in any;
  location: initial ; marked;
end

plant def FEATURE_ATTRIBUTED(alg int x):
  disc bool present in any;
  alg int cost = if present : x else 0 end;
  location: initial ; marked;
end

FM : FEATURE();
FS : FEATURE_ATTRIBUTED(5);
FO : FEATURE();
FR : FEATURE_ATTRIBUTED(5);
FB : FEATURE();
FX : FEATURE_ATTRIBUTED(10);
FE : FEATURE_ATTRIBUTED(5);
FD : FEATURE_ATTRIBUTED(5);
FP : FEATURE_ATTRIBUTED(7);
FC : FEATURE_ATTRIBUTED(5);
FT : FEATURE_ATTRIBUTED(3);

alg bool r1 = FM.present <=> true;
alg bool r2 = FM.present <=> FS.present;
alg bool r3 = FM.present <=> FO.present;
alg bool r4 = FR.present => FM.present;
alg bool r5 = FM.present <=> FB.present;
alg bool r6 = FX.present => FM.present;
alg bool r7 = (FE.present <=> (not(FD.present) and FO.present)) and (FD.present <=> (not(FE.present) and FO.present));
alg bool r8 = FP.present => FB.present;
alg bool r9 = FB.present <=> FC.present;
alg bool r10 = FT.present => FB.present;
alg bool r11 = FP.present => FR.present;
alg bool r12 = not(FD.present and FP.present);

alg bool sys_valid = r1 and r2 and r3 and r4 and r5 and r6 and r7 and r8 and r9 and r10 and r11 and r12;

alg int cost_sum = FS.cost+FR.cost+FX.cost+FE.cost+FD.cost+FP.cost+FC.cost+FT.cost;
alg bool cost_valid = cost_sum <= 30;

plant automaton Validity:
  location: initial sys_valid and cost_valid; marked;
end
\end{lstlisting}
\end{example}

\section{Dynamic configuration}
\label{sec:dynamic}

In the setting discussed in the previous section, the configuration is decided upon initialization of the system and can not change at any later stage. 
In this section we consider the situation that features may configure dynamically.

Different types of reconfiguration can be imagined. 
For example, it can be decided if reconfigurations take place in isolation, or may occur simultaneously. 
We can for example consider the replacement of the euro feature with the dollar feature.
If we do this by first removing the euro feature, and then adding the dollar feature in a next action, these are separate reconfigurations in isolation.
If we replace the euro feature with the dollar feature in a single action, this is simultaneous reconfiguration.
Both alternatives can be modeled in CIF, and are respectively discussed in Sections \ref{subsec:Singlefeature} and \ref{subsec:multifeature}.
The decision on which alternative to use for a model is generally case-specific: can simultaneous reconfiguration practically be achieved, or are reconfiguration actions always performed one-by-one as in single feature configuration?
One can also create a model that contains a combination of simultaneous and single feature reconfiguration.
Nevertheless, the supervisory controller that we will generate in Section \ref{sec:supsynthesis} is always correct-by-construction for the model, and will operate correctly for the physical system if that is adequately represented in the model.

If one allows models to dynamically configure, there may be situations where a specific change in configuration would result in a violation of the feature constraints. 
For the example we mentioned above for reconfiguration in isolation, temporarily neither the euro nor the dollar feature is present.
Hence, it must be decided if such violations of the feature constraints are allowed, this is discussed in Section \ref{subsec:strictness}.

\subsection{Single feature reconfiguration}
\label{subsec:Singlefeature}


In Section \ref{sec:modeling_feature_models}, for each feature an automaton with a variable named \texttt{present} is introduced that captures whether the feature is present.
To allow change of presence status of a feature, the value of the corresponding \texttt{present} variable needs to be able to change.
This can be modeled with a relatively small adaptation to the current feature definition. 
For each feature a \texttt{come} and \texttt{go} event are introduced that represent the addition and removal of the feature from the configuration. 
The resulting feature definition is shown in Listing~\ref{lst:feature_definition}. 
The events \texttt{come} and \texttt{go} are defined inside the automaton. 
As a consequence, there is an instance of both events for each instance of the plant definition. 
These events are referred to by, e.g., \texttt{FM.come} or \texttt{FS.go}.
The same method of adapting the feature automaton definition to allow for dynamic configuration applies for attributed features.

In our examples the come and go events are chosen to be uncontrollable.
This essentially means the supervisory controller safeguards the behavior of the system in the presence of uncontrollable reconfiguration.
When the come and go events are controllable this means the supervisory controller can influence when particular reconfigurations can happen or not dependent on the system state.
It is system dependent whether the come and go events should be modeled controllable or uncontrollable.
It is also possible to have a mix of controllable and uncontrollable reconfiguration in the same model.

\begin{lstlisting}[frame=single,caption={Automaton definition for features with reconfiguration.\label{lst:feature_definition}}]
plant def FEATURE():
  uncontrollable come, go;
  disc bool present in any;
  location: initial; marked;
    edge come when not present do present:=true;
    edge go when present do present:=false;
end
\end{lstlisting}

By instantiating the plant definition from Listing \ref{lst:feature_definition} for all features, a state space is obtained that contains each possible reconfiguration, also those that are invalid by the feature model. 
Restricting reconfiguration to valid configurations is discussed in Section \ref{subsec:strictness}.

\subsection{Multi feature reconfiguration}
\label{subsec:multifeature}
In Section \ref{subsec:Singlefeature} we discussed a situation where features can come and go only one at a time.
In some cases it may be desirable to update the presence of multiple features at a time.
For example, the dollar feature of the coffee machine can be removed and simultaneously the euro feature can be added.
In this way, this reconfiguration can take place without violating the feature constraints, which is impossible with single feature configuration.

To allow updating of the presence variable of multiple feature automata at the same time, a global event can be introduced on which the automata synchronize and their presence variable is updated.
In Listing \ref{lst:multifeature} features \texttt{F1} and \texttt{F2} are alternative features.
The event \texttt{swap12} is declared to define exchanging \texttt{F1} and \texttt{F2}.
By the restriction of the \texttt{Validity} automaton, the system is initially in a valid configuration.
Because of multi feature reconfiguration, \texttt{F1} and \texttt{F2} can be swapped without the system being in an invalid configuration, i.e., always either \texttt{F1} or \texttt{F2} is present.

\begin{lstlisting}[frame=single,caption={Feature automata with multi feature configuration.\label{lst:multifeature}}]
uncontrollable swap12;

plant F1:
  disc bool present in any;
  location: initial ; marked;
    edge swap12 when present do present:=not(present);
end
plant F2:
  disc bool present in any;
  location: initial ; marked;
    edge swap12 when present do present:=not(present);
end

alg bool r1 = (F1.present and not(F2.present)) or (F2.present and not(F1.present));
alg bool sys_valid = r1;

plant automaton Validity:
  location: initial sys_valid; marked;
end

\end{lstlisting}

\subsection{Strictness of the feature constraints}
\label{subsec:strictness}
By allowing features to enter or leave the system, the feature constraints may be violated temporarily.
Two approaches towards the applicability of the feature constraints during reconfiguration are discussed: (1) violation of feature constraints is strictly prohibited, and (2) feature constraints may be violated temporarily.

\paragraph{Strict feature constraints} 
Restricting reconfigurations to valid configurations can be achieved by adding a plant invariant such as presented in Listing \ref{lst:validity_strict}.
Adding this invariant removes all states where \texttt{sys\_valid and cost\_valid} evaluates to \texttt{false}, and all transitions toward these states in the plant's behavior.


\begin{lstlisting}[frame=single,caption={Invariant restricting reconfiguration to valid configurations.\label{lst:validity_strict}}]
plant invariant sys_valid and cost_valid;
\end{lstlisting}

\begin{example}[Reconfiguration with strict feature configuration]
\label{ex:strictexample}
Consider the coffee machine from Example \ref{ex:static}, but now with single feature reconfiguration as discussed in Section \ref{subsec:Singlefeature}, and reconfiguration restricted such that \texttt{sys\_valid} and \texttt{cost\_valid} are always true.
The state space of this system is shown in Figure \ref{fig:CM_statespace}.
Note that this is a single automaton, even though it consists of two unconnected parts.
Just as for the static configuration, there are 16 (initial) states that represent the valid configurations.
Now, switching between configurations is possible by the come and go events.

Since not all states in the state space are connected to each other, we conclude that for some initial configurations, it is not possible to reconfigure to some other configurations.
The seven states on the left hand side are all configurations equipped with the dollar feature, the nine states on the right hand side are all configurations equipped with the euro feature.
In this model, it is impossible to reconfigure from a euro to dollar feature and vice versa. 
This is because for single feature reconfiguration, during reconfiguration either both the euro and dollar feature are present or both are not present, which are invalid configurations.

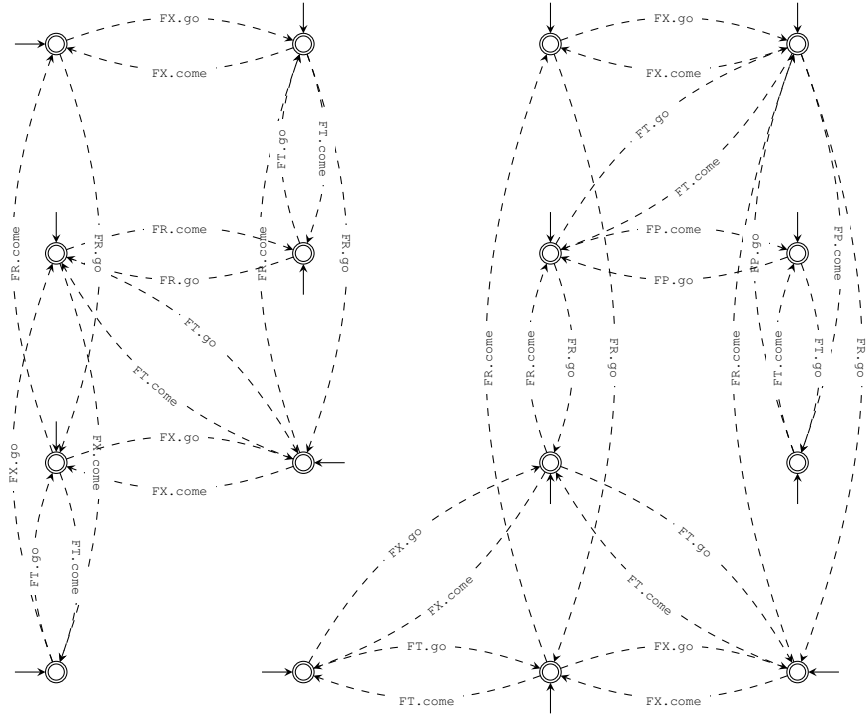
\begin{figure}[htbp]
    \centering
    \begin{tikzpicture}
    \node[state, initial left, accepting] (l15) {};
    \node[state, initial above,right=3.0 of l15,accepting] (l9) {};
    \node[state, initial above,below=2.5 of l15,accepting] (l2) {};
    \node[state, initial below,right=3.0 of l2, accepting] (l10) {};
    \node[state, initial above,below=2.5 of l2,accepting] (l5) {};
    \node[state, initial right,right=3.0 of l5, accepting] (l1) {};
    \node[state, initial left,below=2.5 of l5, accepting] (l6) {};
    
    \node[state, initial above,right=3.0 of l9,accepting] (l16) {};
    \node[state, initial above,right=3.0 of l16, accepting] (l11) {};
    \node[state, initial above,below=2.5 of l16,accepting] (l12) {};
    \node[state, initial above,right=3.0 of l12, accepting] (l14) {};
    \node[state, initial below,below=2.5 of l12,accepting] (l4) {};
    \node[state, initial below,right=3.0 of l4, accepting] (l13) {};
    \node[state, initial below,below=2.5 of l4, accepting] (l7) {};
    \node[state, initial right,right=3.0 of l7,accepting] (l3) {};
    \node[state, initial left,left=3.0 of l7,accepting] (l8) {};

    \draw 
     (l1) edge[sloped,dashed,bend left=20] node[above=-0.2,fill=white]{FR.come} (l9)
     (l1) edge[sloped,dashed,bend left=20] node[above=-0.2,fill=white]{FX.come} (l5)
     (l1) edge[sloped,dashed,bend left=20] node[above=-0.2,fill=white]{FT.come} (l2)
     (l2) edge[sloped,dashed,bend left=20] node[above=-0.2,fill=white]{FR.come} (l10)
     (l2) edge[sloped,dashed,bend left=20] node[above=-0.2,fill=white]{FX.come} (l6)
     (l2) edge[sloped,dashed,bend left=20] node[above=-0.2,fill=white]{FT.go} (l1)
     (l3) edge[sloped,dashed,bend left=20] node[above=-0.2,fill=white]{FR.come} (l11)
     (l3) edge[sloped,dashed,bend left=20] node[above=-0.2,fill=white]{FX.come} (l7)
     (l3) edge[sloped,dashed,bend left=20] node[above=-0.2,fill=white]{FT.come} (l4)
     (l4) edge[sloped,dashed,bend left=20] node[above=-0.2,fill=white]{FR.come} (l12)
     (l4) edge[sloped,dashed,bend left=20] node[above=-0.2,fill=white]{FX.come} (l8)
     (l4) edge[sloped,dashed,bend left=20] node[above=-0.2,fill=white]{FT.go} (l3)
     (l5) edge[sloped,dashed,bend left=20] node[above=-0.2,fill=white]{FR.come} (l15)
     (l5) edge[sloped,dashed,bend left=20] node[above=-0.2,fill=white]{FX.go} (l1)
     (l5) edge[sloped,dashed,bend left=20] node[above=-0.2,fill=white]{FT.come} (l6)
     (l6) edge[sloped,dashed,bend left=20] node[above=-0.2,fill=white]{FX.go} (l2)
     (l6) edge[sloped,dashed,bend left=20] node[above=-0.2,fill=white]{FT.go} (l5)
     (l7) edge[sloped,dashed,bend left=20] node[above=-0.2,fill=white]{FR.come} (l16)
     (l7) edge[sloped,dashed,bend left=20] node[above=-0.2,fill=white]{FX.go} (l3)
     (l7) edge[sloped,dashed,bend left=20] node[above=-0.2,fill=white]{FT.come} (l8)
     (l8) edge[sloped,dashed,bend left=20] node[above=-0.2,fill=white]{FX.go} (l4)
     (l8) edge[sloped,dashed,bend left=20] node[above=-0.2,fill=white]{FT.go} (l7)
     (l9) edge[sloped,dashed,bend left=20] node[above=-0.2,fill=white]{FR.go} (l1)
     (l9) edge[sloped,dashed,bend left=20] node[above=-0.2,fill=white]{FX.come} (l15)
     (l9) edge[sloped,dashed,bend left=20] node[above=-0.2,fill=white]{FT.come} (l10)
    (l10) edge[sloped,dashed,bend left=20] node[above=-0.2,fill=white]{FR.go} (l2)
    (l10) edge[sloped,dashed,bend left=20] node[above=-0.2,fill=white]{FT.go} (l9)
    (l11) edge[sloped,dashed,bend left=20] node[above=-0.2,fill=white]{FR.go} (l3)
    (l11) edge[sloped,dashed,bend left=20] node[above=-0.2,fill=white]{FX.come} (l16)
    (l11) edge[sloped,dashed,bend left=20] node[above=-0.2,fill=white]{FP.come} (l13)
    (l11) edge[sloped,dashed,bend left=20] node[above=-0.2,fill=white]{FT.come} (l12)
    (l12) edge[sloped,dashed,bend left=20] node[above=-0.2,fill=white]{FR.go} (l4)
    (l12) edge[sloped,dashed,bend left=20] node[above=-0.2,fill=white]{FP.come} (l14)
    (l12) edge[sloped,dashed,bend left=20] node[above=-0.2,fill=white]{FT.go} (l11)
    (l13) edge[sloped,dashed,bend left=20] node[above=-0.2,fill=white]{FP.go} (l11)
    (l13) edge[sloped,dashed,bend left=20] node[above=-0.2,fill=white]{FT.come} (l14)
    (l14) edge[sloped,dashed,bend left=20] node[above=-0.2,fill=white]{FP.go} (l12)
    (l14) edge[sloped,dashed,bend left=20] node[above=-0.2,fill=white]{FT.go} (l13)
    (l15) edge[sloped,dashed,bend left=20] node[above=-0.2,fill=white]{FR.go} (l5)
    (l15) edge[sloped,dashed,bend left=20] node[above=-0.2,fill=white]{FX.go} (l9)
    (l16) edge[sloped,dashed,bend left=20] node[above=-0.2,fill=white]{FR.go} (l7)
    (l16) edge[sloped,dashed,bend left=20] node[above=-0.2,fill=white]{FX.go} (l11)
    ;
    \end{tikzpicture}
    \caption{State space of the coffee machine with reconfiguration.}
    \label{fig:CM_statespace}
\end{figure}
\end{example}

\paragraph{Relaxed feature constraints}
It may be desirable to temporarily allow violation of feature constraints during a reconfiguration phase, where the system configuration moves from one valid configuration to another.
This would allow any feature to configure at any moment. 
Consequently, the system may get into a configuration that does not satisfy the feature constraints. 

It should be noted that one may feel the need to express that some of the feature constraints really need to be satisfied at all times. Of course this can still be enforced. 

\begin{example}[Constraints during reconfiguration]
One may allow invalid configurations to be reached.
However, for the coffee machine there may be a constraint when the change feature is present, the coin feature must always be present.
This is achieved using the model fragment from Listing \ref{lst:featureconstraintduring}. 
The resulting state space consists of 1,364 states, among which 16 initial states. There are 13,440 come and go transitions. 
Given the possibilities offered by CIF and the modularly defined feature constraints, it is possible to make more complex constraints.


\begin{lstlisting}[frame=single,caption={Coin feature present when change feature is present.}\label{lst:featureconstraintduring}]
plant invariant FX.present => FO.present;
\end{lstlisting}
\end{example}

\section{Modeling of uncontrolled behavior}
\label{sec:modeling_behavior}

Next, the modeling of uncontrolled behavior of the system components is discussed. 
In Section \ref{sec:requirements}, requirements are defined on this behavior.
This can then be used to synthesize a supervisory controller, which is discussed in Section \ref{sec:supsynthesis}.

\subsection{Behavior of the uncontrolled system}
\label{sec:uncontrolled}

The plant modeling aims at capturing all uncontrolled behavior regardless of features, solely focusing on the potential behavior of the physical components. 

\begin{example}[Coffee machine component behavior]
We model the uncontrolled behavior of the individual components of the coffee machine.
The component-wise behavioral specification of this machine is taken from \cite{Beek2016}.

The system constitutes of the following components: \texttt{Coffee}, \texttt{Tea}, \texttt{Sweet}, \texttt{Ringtone}, \texttt{Coin}, \texttt{Cancel}, and \texttt{Machine}.
For each of the components an automaton is provided that describes its behavior, see Fig. \ref{fig:plants_CM}. 
Although the different models use the same event names (\texttt{done}), because the events are defined within the automata, they are different, and do not synchronize.

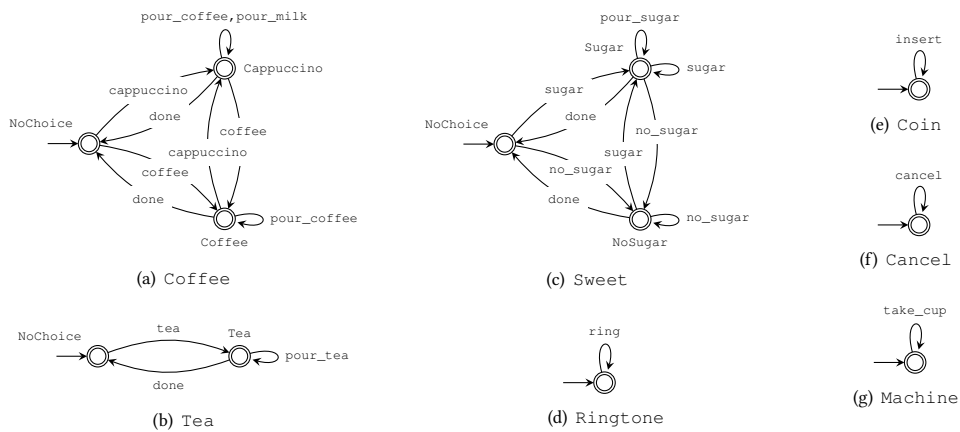
\begin{figure*}[t]
  \begin{minipage}[t][6cm][c]{0.35\columnwidth}
  \centering
     
\subfigure[\texttt{Coffee}]{
    \begin{tikzpicture}
        \node[state, initial, accepting,label={above left:NoChoice}] (NoChoice) {};
        \node[state, accepting, above right=0.8 and 1.6 of NoChoice,label={right:Cappuccino}] (Cappuccino) {};
        \node[state, accepting, below right=0.8 and 1.6 of NoChoice,label={below:Coffee}] (Coffee) {};
        
        \draw   (NoChoice) edge[bend left=20] node[below=-0.1, fill=white]{coffee} (Coffee)
                (NoChoice) edge[bend left=20] node[fill=white]{cappuccino} (Cappuccino)
                (Cappuccino) edge[bend left=20] node[above=-0.1, fill=white]{done} (NoChoice)
                (Cappuccino) edge[bend left=20] node[above right=+0.0 and -0.4, fill=white]{coffee} (Coffee)
                (Coffee) edge[bend left=20] node[above=-0.2, fill=white]{done} (NoChoice)
                (Coffee) edge[bend left=20] node[above=-0.3, fill=white]{cappuccino} (Cappuccino)
                
                (Cappuccino) edge[loop above] node[above=+0.0]{pour\_coffee,pour\_milk} (Cappuccino)
                
                (Coffee) edge[loop right] node[right=+0.0]{pour\_coffee} (Coffee);
    \end{tikzpicture}
     \label{fig:Coffee}}
     
\subfigure[\texttt{Tea}]{
    \begin{tikzpicture}
        \node[state, initial, accepting,label={above left:NoChoice}] (NoChoice) {};
        \node[state, accepting, right= 1.6 of NoChoice,label={above:Tea}] (Tea) {};
        
        \draw   (NoChoice) edge[bend left=20] node[above=+0.0]{tea} (Tea)
                (Tea) edge[bend left=20] node[below=+0.0]{done} (NoChoice)
                
                (Tea) edge[loop right] node[right=+0.0]{pour\_tea} (Tea);
    \end{tikzpicture}
     \label{fig:Tea}}

\end{minipage}
  \begin{minipage}[t][6cm][c]{0.35\columnwidth}
  \centering
  
\subfigure[\texttt{Sweet}]{
    \begin{tikzpicture}
        \node[state, initial, accepting,label={above left:NoChoice}] (NoChoice) {};
        \node[state, accepting, above right=0.8 and 1.6 of NoChoice,label={above left:Sugar}] (Sugar) {};
        \node[state, accepting, below right=0.8 and 1.6 of NoChoice,label={below:NoSugar}] (NoSugar) {};
        
        \draw   (NoChoice) edge[bend left=20] node[below=-0.1, fill=white]{no\_sugar} (NoSugar)
                (NoChoice) edge[bend left=20] node[fill=white]{sugar} (Sugar)
                (Sugar) edge[bend left=20] node[above=-0.1, fill=white]{done} (NoChoice)
                (Sugar) edge[bend left=20] node[above right=+0.0 and -0.4, fill=white]{no\_sugar} (NoSugar)
                (NoSugar) edge[bend left=20] node[above=-0.2, fill=white]{done} (NoChoice)
                (NoSugar) edge[bend left=20] node[above=-0.3, fill=white]{sugar} (Sugar)
                
                (Sugar) edge[loop right] node[right=+0.0]{sugar} (Sugar)
                (Sugar) edge[loop above] node[above=+0.0]{pour\_sugar} (Sugar)
                
                (NoSugar) edge[loop right] node[right=+0.0]{no\_sugar} (NoSugar);
    \end{tikzpicture}
     \label{fig:Sweet}}
     
\subfigure[\texttt{Ringtone}\hspace*{-2em}]{
    \begin{tikzpicture}
        \node[state, initial, accepting] (l0) {};
        
        \draw   (l0) edge[loop above] node[above=+0.0]{ring} (l0);
    \end{tikzpicture}
     \label{fig:Ringtone}}
     
  \end{minipage}
  \begin{minipage}[t][6cm][c]{0.2\columnwidth}
  \centering

\subfigure[\texttt{Coin}]{
    \begin{tikzpicture}
        \node[state, initial, accepting] (l0) {};
        
        \draw   (l0) edge[loop above] node[above=+0.0]{insert} (l0);
    \end{tikzpicture}
     \label{fig:Coin}}
     
\subfigure[\texttt{Cancel}]{
    \begin{tikzpicture}
        \node[state, initial, accepting] (l0) {};
        
        \draw   (l0) edge[loop above] node[above=+0.0]{cancel} (l0);
    \end{tikzpicture}
     \label{fig:Cancel}}
     
\subfigure[\texttt{Machine}]{
    \begin{tikzpicture}
        \node[state, initial, accepting] (l0) {};
        
        \draw   (l0) edge[loop above] node[above=+0.0]{take\_cup} (l0);
    \end{tikzpicture}
     \label{fig:Machine}}
     
\end{minipage}

\caption{Plant automata for the coffee machine.}
\label{fig:plants_CM}
\Description{automata definitions of the coffee machine}
\end{figure*}

The system that is composed of these seven components has a state space of 18 states and 207 transitions, when there is no imposed (supervisory) control, i.e., the events can occur at any time that they are defined in the system.
\end{example}

The CIF model consisting of the component automata does not yet take into account that in specific configurations specific components are not allowed to show behavior, because they are `connected' to features that are not present. 
For example, the event \texttt{ring} of component \texttt{Ringtone} (denoted \texttt{Ringtone.ring}) is only available in case the ringtone feature is part of the configuration. 

In the coffee machine example, for each component there is a one-to-one correspondence with the features. 
In general we require the modeler to indicate for each event that occurs in a component model which features need to be present for that event to be able to occur. 
Note that in some cases the availability of an event may also be dependent on the value of some attribute.
In CIF this can then be captured by means of additional conditions on such events. 
For an event \texttt{e} that requires the presence of feature \texttt{F1} and an attribute value of \texttt{x} for attribute \texttt{A} of feature \texttt{F2}, this is achieved as shown in Listing \ref{lst:link}. 
The connection is stated in the form of a plant, because it models the physical incapability to perform some events when certain features are not present.

\begin{lstlisting}[frame=single,caption={Definition of link between events and features.\label{lst:link}}]
plant automaton event_feature_conditions:
  location: initial; marked;
    edge e when F1.present and F2.A = x;
end
\end{lstlisting}



\begin{example}[Connection between events and features for the coffee machine]
For the coffee machine, the connection between events and features is captured by the plant \texttt{event\_feature\_link} in Listing \ref{lst:event-feature-requirements-CM}.

\begin{lstlisting}[frame=single,caption={Connection between events and features.\label{lst:event-feature-requirements-CM}}]
plant automaton event_feature_link:
  location: initial; marked;
    edge Coffee.cappuccino when FC.present;
    edge Coffee.coffee when FC.present;
    edge Coffee.done when FC.present;
    edge Coffee.pour_coffee when FC.present;
    edge Coffee.pour_milk when FC.present;
    
    edge Tea.done when FT.present;
    edge Tea.pour_tea when FT.present;
    edge Tea.tea when FT.present;
    
    edge Sweet.done when FS.present;
    edge Sweet.no_sugar when FS.present;
    edge Sweet.pour_sugar when FS.present;
    edge Sweet.sugar when FS.present;
    
    edge Ringtone.ring when FR.present;
    
    edge Coin.insert when FO.present;
    
    edge Cancel.cancel when FX.present;
    
    edge Machine.take_cup when FM.present;
end
\end{lstlisting}

\end{example}


The way we expressed the availability of events in relation to the presence of features is conceptually similar to the solution adopted in featured transition systems \cite{Classen2013}. 
In these featured transition systems events are also available conditionally depending on feature presence. 
The most prominent difference between this paper and the approach using featured transition systems is that here the description of the relation between features and events is separated from the behavioral models of the components. 
Another difference is that in the approach using featured transition systems not the uncontrolled system and requirements are modeled, but the supervisory controller is developed directly.

\subsection{Component reappearance}
\label{sec:reappearance}
As a result of reconfiguration, a component may leave or enter the system repeatedly.
Initially, the components are in their initial state as defined in Section \ref{sec:uncontrolled}.
Sometimes it may be required to reinitialize or reset the state of the component when it leaves or enters the system.

Until now, the appearance and disappearance of features is not directly affecting the states of the involved components. 
Therefore, when a feature disappears, and in a future configuration reappears, the components linked with this feature are still in the same state. 

The modeler can easily adapt the plant model such that, for example, a component transitions to some desired reset state whenever the component enters or leaves the system.
For example by adding an edge, labeled with the \texttt{come} or \texttt{go} event of the respective component, from each state in the plant model of the component to its desired reset state.
Because of synchronization, upon occurrence of the \texttt{come} or \texttt{go} event (from the feature plant) the transition with the same label in the component is taken as well. 
When the plant model of the component has an outgoing transition labeled with this reconfiguration transition from each state, the proposed addition does not restrict reconfiguration possibilities.
Otherwise, the reconfiguration event can only take place when the component is in a state where the transition is defined.


\begin{example}[Re-initialization in the coffee machine]
Let us consider the case that we want the tea component to go to the \texttt{NoChoice} location when it leaves the system.
Applying the proposed approach results in the adapted plant automaton shown in Fig. \ref{fig:reentrance}.
If the tea component leaves the system, i.e., event \texttt{FT.go} occurs, when the automaton is in the \texttt{Tea} location, it will transition to \texttt{NoChoice}.
If the event occurs when the automaton is already in \texttt{NoChoice} it will remain there.
Note that in this example the automaton does not use the event \texttt{FT.come} and is therefore not influenced when that particular event occurs.

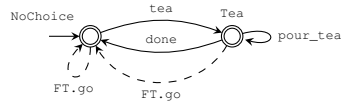
\begin{figure}[htbp]
\centering
\begin{tikzpicture}
        \node[state, initial, accepting,label={above left:NoChoice}] (NoChoice) {};
        \node[state, accepting, right= 1.6 of NoChoice,label={above:Tea}] (Tea) {};
        
        \draw   (NoChoice) edge[bend left=20] node[above=+0.0]{tea} (Tea)
                (Tea) edge[bend left=20] node[above=+0.0]{done} (NoChoice)
                
                (Tea) edge[loop right] node[right=+0.0]{pour\_tea} (Tea)
                
                (Tea) edge[dashed,out=240,in=300,looseness=1.1] node[below=+0.0] {FT.go} (NoChoice)
                (NoChoice) edge[dashed,loop left,looseness=15,out=270,in=225] node[below=+0.0] {FT.go} (NoChoice);
\end{tikzpicture}

    \caption{Adapted plant for resetting the tea component in case of removal from the configuration.}
     \label{fig:reentrance}
     \Description{figure description}
\end{figure}
\end{example}

        


\section{Specification of requirements}
\label{sec:requirements}
In the previous section, we have discussed how to model the uncontrolled system and how to link event occurrences to availability of features. 
In this section, we discuss the modeling of requirements. 
In Section \ref{subsec:behreq} the specification of behavioral requirements is discussed.
In Section \ref{subsec:reqduring} we elaborate on the specification of requirements in the setting of dynamic configuration.

\subsection{Behavioral requirements}
\label{subsec:behreq}
As explained in Section \ref{sec:preliminaries}, requirements are specified by means of automata that synchronize with the plant or by using state-based expressions such as event conditions and state invariants \cite{MarkovskiJBSR10}.
The example below demonstrates several ways to specify the formal requirements, given a set of informal requirements.

\begin{example}[Requirements coffee machine]
Below we state the informal system requirements and their corresponding CIF formulation.
Requirement automata, event conditions, and state invariants are used.
These requirements can be added to the model containing the plant behavior and the feature model.

\begin{enumerate}
\setcounter{enumi}{0}
\item The coffee and tea component can not both be ready to pour:
\end{enumerate}

\noindent
\begin{minipage}[htpb]{\columnwidth}
\vspace*{-1.5\baselineskip}
\begin{lstlisting}[frame=single]
requirement not(Coffee.Coffee and Tea.Tea);
\end{lstlisting}
\end{minipage}

\begin{enumerate}
\setcounter{enumi}{1}
\item Coffee, cappuccino, or tea can only be selected when no choice between them has been made:
\end{enumerate}

\noindent
\begin{minipage}[htpb]{\columnwidth}
\vspace*{-1.5\baselineskip}
\begin{lstlisting}[frame=single]
requirement Coffee.coffee needs Coffee.NoChoice and Tea.NoChoice;
requirement Coffee.cappuccino needs Coffee.NoChoice and Tea.NoChoice;
requirement Tea.tea needs Coffee.NoChoice and Tea.NoChoice;
\end{lstlisting}
\end{minipage}
Note that, in addition to these requirements, these events can only occur when their respective feature is present, because of the connection between the events and feature presence discussed in Section \ref{sec:modeling_behavior}.

\begin{enumerate}
\setcounter{enumi}{2}
\item When the ringtone feature is present, it may only ring (once) after the coffee or tea component is finished:
\end{enumerate}

\noindent
\begin{minipage}[htpb]{\columnwidth}
\vspace*{-1.5\baselineskip}
\begin{lstlisting}[frame=single]
requirement automaton RingAfterBeverageCompletion:
  location NotCompleted:
    initial; marked;
    edge Coffee.done when FR.present goto Completed;
    edge Tea.done when FR.present goto Completed;
    edge Coffee.done, Tea.done when not FR.present;
  location Completed:
    edge Ringtone.ring goto NotCompleted;
end
\end{lstlisting}
\end{minipage}
This requirement shows how the behavioral requirements can be made dependent on the presence of features.
When the ringtone feature is not present, this requirement automaton will stay in \texttt{NotCompleted}.
When the ringtone feature is present, after pouring coffee or tea is done, the automaton needs to transition to the \texttt{Completed} location before the ringtone can be performed.

\begin{enumerate}
\setcounter{enumi}{3}
\item A coin needs to be present to make a selection of beverage and sugar:
\end{enumerate}

\noindent
\begin{minipage}[htpb]{\columnwidth}
\vspace*{-1.5\baselineskip}
\begin{lstlisting}[frame=single]
plant automaton CoinPresence:
  monitor;
  location NoCoinPresent:
    initial; marked;
    edge Coin.insert goto CoinPresent;
  location CoinPresent:
    edge Cancel.cancel goto NoCoinPresent;
    edge Machine.take_cup goto NoCoinPresent;
end

requirement Coffee.coffee needs CoinPresence.CoinPresent;
requirement Coffee.cappuccino needs CoinPresence.CoinPresent;
requirement Tea.tea needs CoinPresence.CoinPresent;
requirement Sweet.sugar needs CoinPresence.CoinPresent;
requirement Sweet.no_sugar needs CoinPresence.CoinPresent;
\end{lstlisting}
\end{minipage}
This requirement showcases the use of a \texttt{monitor} automaton.
In literature this is sometimes also called an observer automaton.
The automaton \texttt{CoinPresence} never disables any event, but simply tracks whether a coin is present in the system.
The state of \texttt{CoinPresence} is used in the subsequent event conditions so that only a selection can be made when a coin is present.

\begin{enumerate}
\setcounter{enumi}{4}
\item Coffee is only poured once:
\end{enumerate}

\noindent
\begin{minipage}[htpb]{\columnwidth}
\vspace*{-1.5\baselineskip}
\begin{lstlisting}[frame=single]
plant automaton CoffeePoured:
  monitor;
  location NotPoured:
    initial;marked;
    edge Coffee.pour_coffee goto Poured;
  location Poured:
    edge Machine.take_cup goto NotPoured;
end

requirement Coffee.pour_coffee needs CoffeePoured.NotPoured;
\end{lstlisting}
\end{minipage}
Once more, a monitor automaton is used. 
This automaton tracks whether Coffee has been poured.
Note that we have similar informal requirements with CIF formalizations for only pouring tea and milk once; resulting in automata \texttt{TeaPoured} and \texttt{MilkPoured}.

\begin{enumerate}
\setcounter{enumi}{5}
\item Coffee and tea should not be mixed, and tea and milk should not be mixed:
\end{enumerate}

\noindent
\begin{minipage}[htpb]{\columnwidth}
\vspace*{-1.5\baselineskip}
\begin{lstlisting}[frame=single]
requirement not(CoffeePoured.Poured and TeaPoured.Poured);
requirement not(TeaPoured.Poured and MilkPoured.Poured);
\end{lstlisting}
\end{minipage}
Here the previously defined states of the monitor automata are used in additional state invariants.

\begin{enumerate}
\setcounter{enumi}{6}
\item When coffee is selected, the coffee component is done after pouring only coffee. When cappuccino is selected, the coffee component is done after pouring both coffee and milk:
\end{enumerate}

\noindent
\begin{minipage}[htpb]{\columnwidth}
\vspace*{-1.5\baselineskip}
\begin{lstlisting}[frame=single]
requirement Coffee.done needs (Coffee.Coffee and CoffeePoured.Poured) or (Coffee.Cappuccino and CoffeePoured.Poured and MilkPoured.Poured);
\end{lstlisting}
\end{minipage}

\begin{enumerate}
\setcounter{enumi}{7}
\item When sugar is selected, it is poured twice:
\end{enumerate}

\noindent
\begin{minipage}[htpb]{\columnwidth}
\vspace*{-1.5\baselineskip}
\begin{lstlisting}[frame=single]
requirement automaton PourSugarTwice:
  disc int[0..2] count=0;
  location Idle:
    initial; marked;
    edge Sweet.sugar goto SugarNeeded;
    edge Sweet.done when Sweet.NoSugar;
    edge Machine.take_cup do count:=0;
  location SugarNeeded:
    edge Sweet.pour_sugar when count<2 do count:=count+1;
    edge Sweet.done when count=2 goto Idle;
    edge Machine.take_cup do count:=0;
end
\end{lstlisting}
\end{minipage}
Here we use a requirement automaton that is an extended finite automaton to keep track of a count of how many times sugar has been poured.
Pouring sugar is only finished after it has been performed twice, and the counter is reset to zero when the cup is taken from the system.

\begin{enumerate}
\setcounter{enumi}{7}
\item Cancellation is only possible before anything has been poured:
\end{enumerate}

\noindent
\begin{minipage}[htpb]{\columnwidth}
\vspace*{-1.5\baselineskip}
\begin{lstlisting}[frame=single]
requirement Cancel.cancel needs CoffeePoured.NotPoured and TeaPoured.NotPoured and MilkPoured.NotPoured and PourSugarTwice.count=0;
\end{lstlisting}
\end{minipage}

\begin{enumerate}
\setcounter{enumi}{8}
\item The cup can only be taken from the machine when the pouring of coffee, tea, and sugar is done and no new selection was made:
\end{enumerate}

\noindent
\begin{minipage}[htpb]{\columnwidth}
\vspace*{-1.5\baselineskip}
\begin{lstlisting}[frame=single]
requirement automaton TakeCupWhenCoffeeOrTeaDone:
  location NotPoured:
    initial;marked;
    edge Coffee.done goto Done;
    edge Tea.done goto Done;
  location Done:
    edge Machine.take_cup goto NotPoured;
end

requirement automaton TakeCupWhenSugarDone:
  location NotPoured:
    initial;marked;
    edge Sweet.done goto Done;
  location Done:
    edge Machine.take_cup goto NotPoured;
end

requirement Machine.take_cup needs Coffee.NoChoice and Sweet.NoChoice and Tea.NoChoice;
\end{lstlisting}
\end{minipage}

\end{example}

\subsection{Requirements during configuration}
\label{subsec:reqduring}
As discussed in Section \ref{subsec:strictness}, it may be beneficial to allow invalid configurations so that reconfiguration can take place.
In this case, decisions must be made about allowed behavior in such configurations.
There are several ways to deal with the specification of allowed behavior during the configuration phase:
(1) disable some events from occurring, and (2) additional requirements.
Each of these approaches may be suitable for certain applications.
In the following subsections these possibilities are investigated. 

\subsubsection{Disabling events}
An approach to restrict the behavior during reconfiguration is to disable some events in case the system is in an invalid configuration.
For each such an event, an event condition, such as the one presented for event \texttt{e} in Listing \ref{lst:disable_event}, can be defined that restricts that event to occur only when the system is in a valid configuration. 

\noindent
\begin{minipage}[htpb]{\columnwidth}
\begin{lstlisting}[frame=single,caption={Disabling an event in an invalid system configuration.\label{lst:disable_event}}]
requirement e needs sys_valid;
\end{lstlisting}
\end{minipage}

This approach assumes that the system will exhibit safe behavior by not exercising any of the events disabled in this way. 
As soon as the system returns to a valid configuration these events are no longer disabled.
Such event disablement requirements can also be expressed for particular configurations, as discussed in the next example.

\begin{example}[Event disablement in coffee machine during reconfiguration]
In the coffee machine, invalid configurations may sometimes be allowed. 
For example, exchanging the euro and dollar feature through two subsequent single feature reconfiguration events is achieved by either having none or both features present during the reconfiguration.
Let us consider the case that both the euro and dollar feature are present.
It may be unsafe to cancel the order in this situation, as it is unclear from which feature the coin should be returned.
This can be avoided by adding a requirement as given in Listing \ref{lst:reqduring}.

\begin{lstlisting}[frame=single,caption={Constraint during invalid configuration.\label{lst:reqduring}}]
requirement Cancel.cancel needs not(FE.present and FD.present);
\end{lstlisting}
\end{example}

\subsubsection{Additional requirements}
Another approach is stating additional requirements for the transitional situation. 
For the coffee machine these are detailed in the next example.

\begin{example}[Dynamic configuration constraints for the coffee machine]
We consider the situation that it is allowed that the sweet feature is not present during reconfiguration.
However, in case the sweet feature is currently ready to pour sugar, it needs to always be present.
Listing \ref{lst:reqduring2} shows a formalization of this requirement, that restricts the possible (invalid) configurations that can be reached depending on the current state of the components.
\begin{lstlisting}[frame=single,caption={Constraint during invalid configuration.\label{lst:reqduring2}}]
requirement Sweet.Sugar => FS.present;
\end{lstlisting}
\end{example}

\section{Supervisory controller synthesis}
\label{sec:supsynthesis}
The automata model obtained from the feature model in Section \ref{sec:dynamic}, the uncontrolled behavior specification from Section \ref{sec:modeling_behavior}, and the behavioral requirements from Section \ref{sec:requirements} can be placed into a single model.
Next, supervisory control synthesis \cite{Ramadge1987} can be applied.
A supervisory controller is generated as an automaton that controls the system through synchronizing events.
The system under control consists of the supervisor synchronized with all plant and requirement automata.
By construction of the supervisor, the system under control is:
\begin{enumerate}
    \item \textit{Safe}: the requirements are always adhered to. If in synchronous composition the plant can execute an event, but it is not possible in a requirement automaton that synchronizes on that event, the event is prevented from occurring.
    \item \textit{Nonblocking}: a marked state can always be reached. Since we modeled the system using multiple automata, this means in the controlled behavior it is always possible to follow a sequence of events such that all automata are simultaneously in a marked state. For example, if we consider the \texttt{CoinPresence} requirement of Section \ref{subsec:behreq}, we are sure that the system can always return to the state that no coins are present.
    \item \textit{Controllable}: the supervisor does not disallow uncontrollable events to occur. In our model of the coffee machine all reconfiguration actions are uncontrollable. The supervisor always allows these to happen, and prevents the sytem from reaching states that are not allowed in a particular configuration, if we may uncontrollably reconfigure to that configuration.
    \item \textit{Maximally permissive}: no behavior is disabled that does not strictly need to be disallowed to satisfy the aforementioned properties. This makes sure the supervisory controller does not restrict any behavior that is perfectly fine to occur.
\end{enumerate}
Note that a single supervisory controller is generated that applies to all system configurations.

\begin{example}[Supervisory controller synthesis for the coffee machine]
We consider the coffee machine with dynamic single feature configuration, where invalid configurations are never allowed, and the requirements of Section \ref{sec:requirements} are applied.
Applying supervisory controller synthesis to the described system results in a supervisory controller that applies the guards formulated in Listing \ref{lst:supervisor} to the controllable events. 
Note that the guard for \texttt{Cancel.cancel} is not displayed because the expression is very long.
The state space of the system under control contains 6,240 states and 35,336 transitions. 

\begin{lstlisting}[frame=single,caption={Additional guards provided by supervisory controller synthesis.\label{lst:supervisor}}]
supervisor automaton sup:
  alphabet Coin.insert, Cancel.cancel, Sweet.sugar, Sweet.no_sugar, Sweet.pour_sugar, Sweet.done, Ringtone.ring, Coffee.cappuccino, Coffee.coffee, Coffee.pour_coffee, Coffee.pour_milk, Coffee.done, Tea.tea, Tea.pour_tea, Tea.done, Machine.take_cup;
  location:
    initial;
    marked;
    edge Cancel.cancel when ...;
    edge Coffee.cappuccino when CoinPresence.CoinPresent and not Coffee.Coffee and (Tea.NoChoice and TakeCupWhenCoffeeOrTeaDone.NotPoured);
    edge Coffee.coffee when CoinPresence.CoinPresent and not Coffee.Cappuccino and (Tea.NoChoice and TakeCupWhenCoffeeOrTeaDone.NotPoured);
    edge Coffee.done when not Coffee.Cappuccino and CoffeePoured.Poured or Coffee.Cappuccino and (CoffeePoured.Poured and MilkPoured.Poured);
    edge Coffee.pour_coffee when CoffeePoured.NotPoured;
    edge Coffee.pour_milk when MilkPoured.NotPoured;
    edge Coin.insert when true;
    edge Machine.take_cup when true;
    edge Ringtone.ring when true;
    edge Sweet.done when true;
    edge Sweet.no_sugar when CoinPresence.CoinPresent and not Sweet.Sugar and (PourSugarTwice.Idle and TakeCupWhenSugarDone.NotPoured) or (CoinPresence.CoinPresent and (not Sweet.Sugar and PourSugarTwice.SugarNeeded) or CoinPresence.CoinPresent and (Sweet.Sugar and PourSugarTwice.count = 2));
    edge Sweet.pour_sugar when true;
    edge Sweet.sugar when CoinPresence.CoinPresent and TakeCupWhenSugarDone.NotPoured;
    edge Tea.done when true;
    edge Tea.pour_tea when TeaPoured.NotPoured;
    edge Tea.tea when CoinPresence.CoinPresent and (Coffee.NoChoice and TakeCupWhenCoffeeOrTeaDone.NotPoured);
end
\end{lstlisting}
\end{example}

\section{Case study: Body Comfort System}
\label{sec:usecase}
\begin{figure}[htbp]
\centerline{
\includegraphics[width=\columnwidth]{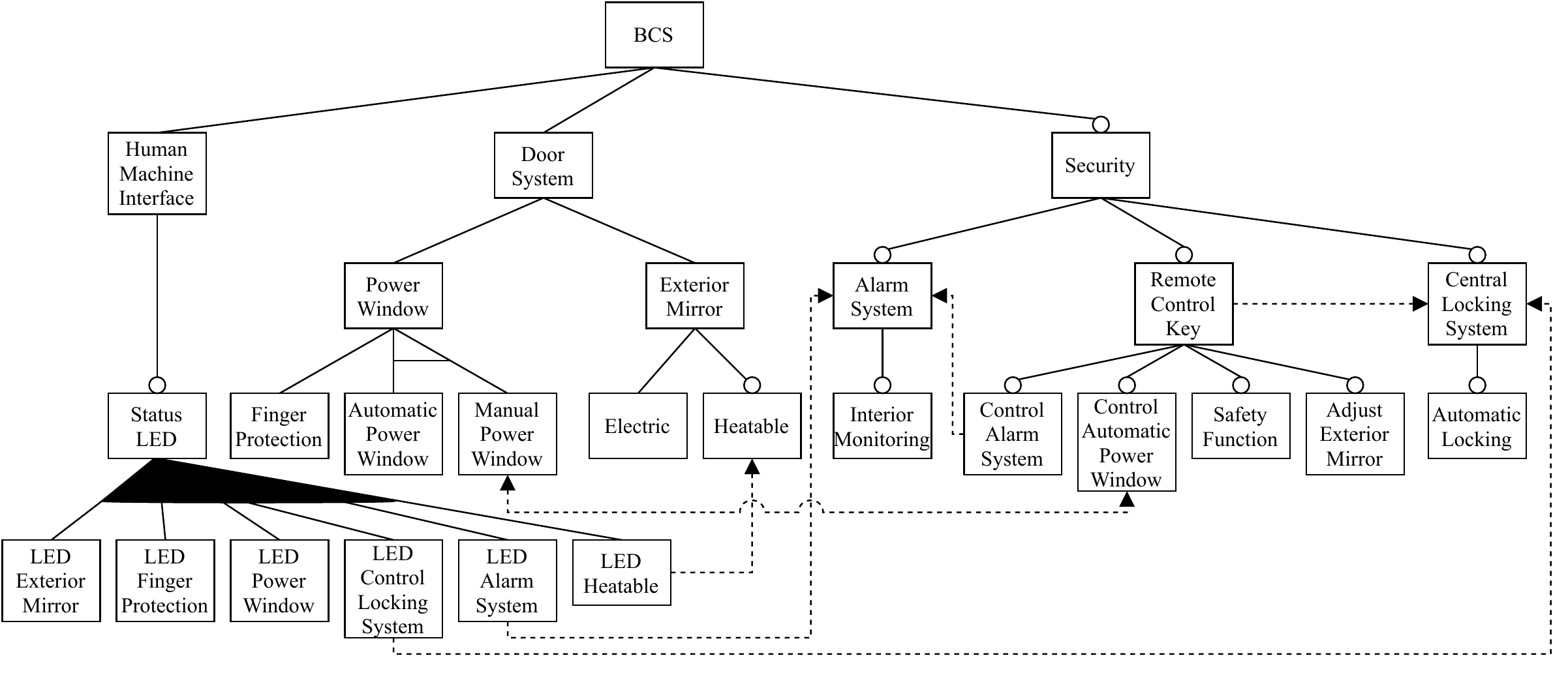}
}
\caption{Feature model of the Body Comfort System \cite{Lity2013}.}
\label{fig:BCS_FM}
\end{figure}

We present a case study on the Body Comfort System (BCS) \cite{Lity2013}, which is a frequently used benchmark in (S)PLE-related literature \cite{Lochau2014,Fragal2016,Lity2017,Lachmann2016}. 
It is a product line originating from the automotive industry. 
It contains a number of standard and optional features, such as LED's in the human machine interface, manual or automatic windows, security options such as an alarm system, and more. 
The feature model of the BCS is given in Figure \ref{fig:BCS_FM}.
This feature model allows for 11,616 valid configurations.
Note that for this case study, only features without attributes are considered.

The feature model is modeled in CIF as outlined in Section \ref{sec:modeling_feature_models}, this results in Listing \ref{lst:BCS_FM_CIF}.

\begin{lstlisting}[frame=single,caption={BCS feature model in CIF.\label{lst:BCS_FM_CIF}}]
plant def FEATURE():
  uncontrollable come,go;
  disc bool present in any;
  location: initial ; marked;
     edge come when not present do present:=true;
     edge go when present do present:=false;
end

// Feature declaration by level in FM
// Level 1
FBCS:FEATURE();
// Level 2
FHMI:FEATURE(); FDoor:FEATURE(); FSecu:FEATURE();
// Level 3
FPowerW:FEATURE(); FMir:FEATURE(); FAlarm:FEATURE();FRCKey:FEATURE();FCLS:FEATURE();
//Level 4
FLED:FEATURE(); FFingerP:FEATURE(); FAutoPW:FEATURE(); FManPW:FEATURE(); FMirE:FEATURE(); FMirHeat:FEATURE(); FInterMon:FEATURE(); FCtrAlarm:FEATURE(); FCtrAutoPW:FEATURE(); FSafe:FEATURE(); FAdjMir:FEATURE(); FAutoL:FEATURE();
//Level 5
FLEDMir:FEATURE(); FLEDFP:FEATURE(); FLEDPW:FEATURE(); FLEDCLS:FEATURE(); FLEDAlarm:FEATURE(); FLEDHeat:FEATURE();

//Feature relations
// Level 1
alg bool r11 = FBCS.present; // Root feature present
// Level 2
alg bool r21 = FBCS.present <=> FHMI.present; // HMI mandatory
alg bool r22 = FBCS.present <=> FDoor.present; // Door mandatory
alg bool r23 = FSecu.present => FBCS.present; // Security optional
// Level 3
alg bool r31 = FDoor.present <=> FPowerW.present; // PW mandatory
alg bool r32 = FDoor.present <=> FMir.present; //EM mandatory
alg bool r33 = FAlarm.present => FSecu.present; // AS optional
alg bool r34 = FCLS.present => FSecu.present; // CLS optional
alg bool r35 = FRCKey.present => FSecu.present; // RCK optional
// Level 4
alg bool r41 = FLED.present => FHMI.present; // LED optional
alg bool r42 = (FManPW.present <=> (not FAutoPW.present and FPowerW.present)) and
                      (FAutoPW.present <=> (not FManPW.present and FPowerW.present)); //Manual or automatic PW
alg bool r43 = FPowerW.present <=> FFingerP.present; // Finger Protection mandatory
alg bool r44 = FMir.present <=> FMirE.present; // Electric exterior mirror mandatory
alg bool r45 = FMirHeat.present => FMir.present; // Mirror heating optional
alg bool r46 = FInterMon.present => FAlarm.present; // Interior monitoring optional
alg bool r47 = FCtrAlarm.present => FRCKey.present; // Control alarm optional
alg bool r48 = FCtrAutoPW.present => FRCKey.present; // Control automatic power window optional
alg bool r49 = FSafe.present => FRCKey.present; // Safety optional
alg bool r410 = FAdjMir.present => FRCKey.present; // Adjust exterior mirror optional
alg bool r411 = FAutoL.present => FCLS.present; // Automatic locking optional
//Level 5
alg bool r51 = FLED.present <=> (FLEDAlarm.present or FLEDFP.present or FLEDCLS.present or FLEDPW.present or FLEDMir.present or FLEDHeat.present);
// cross tree relations
alg bool rx1 = FLEDAlarm.present => FAlarm.present; //LED alarm requires Alarm
alg bool rx2 = FLEDCLS.present => FCLS.present; //LED central requires central locking
alg bool rx3 = FLEDHeat.present => FMirHeat.present; //LED heat mirror requires heated mirror
alg bool rx4 = not(FManPW.present and FCtrAutoPW.present); //Manual power windows excludes control autoPW
alg bool rx5 = FCtrAlarm.present => FAlarm.present; //Control alarm requires Alarm system
alg bool rx6 = FRCKey.present => FCLS.present; //Remote control key requires central locking system

alg bool sys_valid = r11 and r21 and r22 and r23 and r31 and r32 and r33 and r34 and r35 and r41 and r42 and r43 and r44 and r45 and r46 and r47 and r48 and r49 and r410 and r411 and r51 and rx1 and rx2 and rx3 and rx4 and rx5 and rx6;

plant automaton Validity:
  location: initial sys_valid; marked;
end
\end{lstlisting}

In this model, the system is initially in a valid configuration, but can reconfigure to any configuration through the come and go events. 
I.e., all invalid configurations are allowed during reconfiguration.
This model (without any component behavior added) has 134,217,728 reachable states.
Of these states, 11,616 states are initial states representing the valid configurations.

In \cite{Lity2013} state machine test models are constructed for product instances of the BCS.
In \cite{Tuitert} a component wise behavioral model of the BCS is made, as an interpretation of the models in \cite{Lity2013}.
We will use the models from \cite{Tuitert} here.

As an example, we consider the alarm system and interior monitoring features.
The uncontrolled behavior of the relevant components is defined by the automata shown in Listing \ref{lst:BCS_uncontrolled}.
In the event names, "\texttt{c\_}" and "\texttt{u\_}" are used as prefixes to respectively indicate controllable or uncontrollable events.

\begin{lstlisting}[frame=single,caption={BCS uncontrolled behavior of the alarm system and interior monitoring.\label{lst:BCS_uncontrolled}}]
plant automaton AlarmSystem:
  controllable c_on, c_off, c_deactivated, c_activated, c_IM_detected;
  uncontrollable u_detected, u_time_elapsed;
  location Deactivated:
    edge c_activated goto Activated;
  location Activated:
    initial; marked;
    edge c_on goto On;
    edge c_deactivated goto Deactivated;
  location On:
    edge c_off goto Activated;
    edge u_detected goto Alarm_detected;
    edge c_IM_detected goto Alarm_detected;
  location Alarm_detected:
    edge c_off goto Activated;
    edge u_time_elapsed goto On;
end

plant automaton InteriorMonitoring:
  uncontrollable u_detected, u_clear;
  controllable c_on, c_off;
  location Off:
    initial; marked;
    edge c_on goto On;
  location On:
    edge c_off goto Off;
    edge u_detected goto Detected;
  location Detected:
    edge u_clear goto On;
    edge c_off goto Off;
end
\end{lstlisting}

The events of the automata are linked to their presence in Listing \ref{lst:BCS_presence}.
One can observe that the events of \texttt{AlarmSystem} and \texttt{InteriorMonitoring} are dependent on the presence features \texttt{FAlarm} and \texttt{FInterMon} respectively.
However, the event \texttt{AlarmSystem.c\_IM\_detected} requires both features to be present.

\begin{lstlisting}[frame=single,caption={BCS presence check alarm system.\label{lst:BCS_presence}}]
plant automaton PRESENCE_UNCONTROLLED_AS:
  location: initial; marked;
    edge AlarmSystem.u_detected when FAlarm.present;
    edge AlarmSystem.u_time_elapsed when FAlarm.present;
    edge AlarmSystem.c_on when FAlarm.present;
    edge AlarmSystem.c_off when FAlarm.present;
    edge AlarmSystem.c_deactivated when FAlarm.present;
    edge AlarmSystem.c_IM_detected when FAlarm.present and FInterMon.present;
    edge InteriorMonitoring.u_detected when FInterMon.present;
    edge InteriorMonitoring.u_clear when FInterMon.present;
    edge InteriorMonitoring.c_on when FInterMon.present;
    edge InteriorMonitoring.c_off when FInterMon.present;
end
\end{lstlisting}

Requirements for these components are given in Listing \ref{lst:BCS_reqs}.
One can see how these refer to other relevant components in the system such as \texttt{Key\_lock} and \texttt{RCK\_CLS}, which are the physical key and the remote control key of the central locking system.

\begin{lstlisting}[frame=single,caption={BCS requirements alarm system.\label{lst:BCS_reqs}}]
requirement AlarmSystem.c_on needs Key_lock.Locked or RCK_CLS.Locked;
requirement AlarmSystem.c_off needs Key_lock.Unlocked or RCK_CLS.Unlocked;
requirement AlarmSystem.c_deactivated needs Key_lock.Unlocked or RCK_CLS.Unlocked;
requirement AlarmSystem.c_IM_detected needs InteriorMonitoring.Detected;
requirement InteriorMonitoring.c_off needs Key_lock.Unlocked or RCK_CLS.Unlocked;
\end{lstlisting}

In the complete model\footnote{Note that also all complete CIF models we use of the BCS are available here: \url{https://github.com/sbthuijsman/TECS_PLE}}, there are in total 31 plant automata representing the behavior of the components.
Additionally, there are 27 feature automata and 18 plant automata that link the component events to the presence of the features.
55 requirements are specified. 
Only event condition requirements are used.
Using CIF, synthesis can successfully be applied to this system.

Some relevant state space sizes are mentioned in Table \ref{tab:ss_BCS}.
The worst-case state space is calculated by calculating the product of the number of states in each automaton.
The other state space sizes all denote the reachable states from the initial states.
For the uncontrolled systems, no synthesis or requirements are applied yet.
For the controlled systems, the state space size is the number of reachable states in the system controlled by the synthesized supervisory controller.
Using a standard personal computer and applying supervisory controller synthesis in CIF using the default settings, the supervisor is obtained in roughly 0.3 seconds for each of the static and the dynamic case.
For these computations, the CIF application requires no more than 0.5 GB of memory.
Even though a state space in the order of $10^{20}$ can be considered large, CIF has been shown capable of performing supervisory controller synthesis for systems with much larger state spaces \cite{Reijnen2020}.
It should be noted that, since CIF uses symbolic supervisory controller synthesis using Binary Decision Diagrams (BDDs), the computational effort of synthesis is dependent on more factors than state space size \cite{Thuijsman2019}.
Supervisory controller synthesis for the static case requires 19,614 peak used BDD nodes and 1,864,598 BDD operations.
For the dynamic case this is 26,140 peak used BDD nodes and 2,039,318 operations.
These BDD-based metrics of computational effort of supervisory controller synthesis are detailed in \cite{Thuijsman2019} and measurements for benchmark systems are provided in the same work.

\begin{table}[th]
\caption{State space sizes BCS.}
\label{tab:ss_BCS}
\begin{tabular}{ll}
\textbf{State space} & \textbf{Number of states} \\ \hline
Worst-case           & $7.7\cdot 10^{20}$        \\
Uncontrolled static  & $3.2\cdot 10^{14}$        \\
Uncontrolled dynamic & $6.2\cdot 10^{20}$        \\
Controlled static    & $7.6\cdot 10^{13}$        \\
Controlled dynamic   & $1.1\cdot 10^{20}$       
\end{tabular}
\end{table}

Even though the BCS is a frequently used benchmark in literature related to product line engineering \cite{Lity2013,Lochau2014,Fragal2016,Lity2017,Lachmann2016}, as far as we are aware there is no existing work to which we can compare these results.
In fact, only in \cite{Tuitert} the first models of the BCS were made that were suitable for application of supervisory control theory.

With the BCS use case, we have shown that CIF is capable of both modeling an industrial sized product line through the use of feature models, and synthesizing a supervisory controller for this system in which dynamic reconfiguration is allowed.



\section{Concluding Remarks}
\label{sec:conclusions}

We have presented a framework for engineering supervisory controllers for product lines of which the valid configurations are described by a feature model and where dynamic configuration of the features is allowed.
The CIF language has shown to be adequate for modeling the involved concepts.
It was shown how the presence and absence of features can be modeled, and how this presence can update through single or multi feature configuration.
Component wise modeling of the system behavior has been demonstrated, where the presence of features influences the possible behavior.
It was shown how requirements can be formulated that take the presence of the features into account, and how requirements can be strengthened when the system is in an invalid configuration.
Supervisory controller synthesis can be applied such that the maximal nonblocking, controllable, and safe behavior under control is obtained.
The method was demonstrated using the coffee machine as a running example.
Although the coffee machine system is small, feasibility of industrial application has been demonstrated with the much larger BCS use case.

\begin{acks}
Research leading to these results has received funding from the EU ECSEL Joint Undertaking under grant agreement n\textsuperscript{o} 826452 (project Arrowhead Tools) and from the partners national programs/funding authorities.
\end{acks}

\bibliographystyle{ACM-Reference-Format}
\bibliography{references_fixed}

\end{document}